\documentclass[useAMS,usenatbib,twocolumn]{mn2e}
\usepackage{subfigure}
\usepackage{amsmath}
\usepackage{amssymb}
\usepackage{rotating}
\usepackage{times}
\usepackage{graphicx}
\usepackage{lscape}
\usepackage{xcolor}
\def\etal{{\it et al.}}

\title[GRB magnetar central engines and the LT correlation]{Constraining properties of GRB magnetar central engines using the observed plateau luminosity and duration correlation}
\author[A. Rowlinson \etal]{A. Rowlinson$^{1,2}$\thanks{E-mail:Antonia.Rowlinson@csiro.au}, B.~P. Gompertz$^{3}$, M. Dainotti$^{4,5,6}$, P.~T. O'Brien$^{3}$, R.~A.~M.~J. Wijers$^{2}$, \newauthor A. J. van der Horst$^{2}$ \\
$^{1}$CSIRO Astronomy and Space Science, PO Box 76, Epping, NSW 1710, Australia\\
$^{2}$Anton Pannekoek Institute, University of Amsterdam, Postbus 94249, 1090 GE Amsterdam, The Netherlands\\
$^{3}$Department of Physics \& Astronomy, University of Leicester, University Road, Leicester, LE1 7RH, UK\\
$^{4}$Astrophysical Big Bang Laboratory, 2-1 Hirosawa, Wako, Saitama 351-0198, Japan \\
$^{5}$Astronomy Department, Stanford University, 382 Via Pueblo Mall, Stanford CA 94305-4060, California \\
$^{6}$Astronomical Observatory, Jagiellonian University, ul. Orla 171, 30-244 Cracow, Poland
}

\begin{document}

\pagerange{\pageref{firstpage}--\pageref{lastpage}} \pubyear{000}
\maketitle            

\label{firstpage}

\begin{abstract}

An intrinsic correlation has been identified between the luminosity and duration of plateaus in the X-ray afterglows of Gamma-Ray Bursts \citep[GRBs;][]{dainotti2008}, suggesting a central engine origin. The magnetar central engine model predicts an observable plateau phase, with plateau durations and luminosities being determined by the magnetic fields and spin periods of the newly formed magnetar. This paper analytically shows that the magnetar central engine model can explain, within the $1\sigma$ uncertainties, the correlation between plateau luminosity and duration. The observed scatter in the correlation most likely originates in the spread of initial spin periods of the newly formed magnetar and provides an estimate of the maximum spin period of $\sim35$ ms (assuming a constant mass, efficiency and beaming across the GRB sample). Additionally, by combining the observed data and simulations, we show that the magnetar emission is most likely narrowly beamed and has $\lesssim20$\% efficiency in conversion of rotational energy from the magnetar into the observed plateau luminosity. The beaming angles and efficiencies obtained by this method are fully consistent with both predicted and observed values. We find that Short GRBs and Short GRBs with Extended Emission lie on the same correlation but are statistically inconsistent with being drawn from the same distribution as Long GRBs, this is consistent with them having a wider beaming angle than Long GRBs. 

\end{abstract}

\begin{keywords}
Gamma-Ray Bursts, Magnetars
\end{keywords}

\section{Introduction}

Gamma-Ray Bursts (GRBs) show a great diversity in their prompt emission lightcurves and a broad range of isotropic energies ($E_{\rm iso}$), whilst sharing common features in their X-ray afterglow lightcurves leading to the identification of a canonical X-ray lightcurve \citep{obrien2006, nousek2006,zhang2006}. In 42\% of X-ray afterglows we observe a plateau phase, a shallow decay regime in the lightcurve, which is evidence of ongoing energy injection from the central engine \citep{evans2009}. By fitting the rest-frame afterglows of GRBs with an analytic model representing the common features of GRBs \citep{willingale2007}, \cite{dainotti2008} discovered a formal (i.e. robust against standard statistical correlation tests) anti-correlation between the luminosity and the rest-frame duration of the plateau phase (hereafter referred to as the LT correlation). In subsequent analysis, this correlation has been shown to be independent of underlying cosmological or systematic effects \citep{dainotti2010,dainotti2011,dainotti2013} and suggested as a tool for constraining cosmological parameters \citep{cardone2009,cardone2010,dainotti2013b,postnikov2014}. Understanding the physical origin of this correlation will provide significant clues about the central engine activity.

GRBs are thought to originate from at least two progenitors; long GRBs (LGRBs) associated with the collapse of a massive star \citep[e.g.][]{woosley1993} and short GRBs (SGRBs) originating from the merger of two neutron stars (NSs) or a NS and a black hole \citep[BH;][]{lattimer1976, eichler1989,narayan1992}. Hence, in addition to being able to explain both the prompt emission and late-time energy injection, the central engine needs to be able to form via different mechanisms. Two main contenders have been proposed for the central engine: a BH \citep[e.g.][]{woosley1993} or a newly formed, highly magnetised, millisecond NS \citep[magnetar; e.g.][]{usov1992, zhang2001, metzger2010}. There is an increasingly large sample of GRBs that are well fit with the magnetar model, including SGRBs \citep[where the energy injection is difficult to explain using a BH central engine, e.g.][]{rowlinson2010,rowlinson2013,postigo2013,lu2014}, short GRBs with extended emission \citep[EE SGRBs, e.g.][]{gompertz2013,gompertz2013b}, and some LGRBs \citep[e.g.][]{troja2007, lyons2010, dallosso2011, bernardini2012, bernardini2013, lu2014,yi2014}. 

Using the results from 12 LGRBs fitted with the magnetar central engine model, \cite{bernardini2012} showed that the magnetar central engine model is able to reproduce the LT correlation observed by \cite{dainotti2008}, assuming that the magnetar is injecting energy into the forward shock. However, there are a number of candidate magnetar plateaus that exhibit very steep decay phases that are inconsistent with energy injection into the forward shock \citep{troja2007,lyons2010,rowlinson2010,rowlinson2013} and other emission mechanisms require further investigation, such as energy injection into the reverse shock \citep{leventis2014,eerten2014,eerten2014b}. \cite{bernardini2012} randomly select magnetar parameters from Gaussian distributions representing the fitted parameters from their sample of 12 LGRBs. These fitted magnetar parameters may not be representative of the total sample of possible GRB magnetar central engines due to observational selection effects.

In this paper, we explore a magnetar central engine as the physical origin of the LT correlation and show that it is intrinsic to the standard theory of the spindown magnetar wind, irrespective of the emission mechanism. Section 2 extends the work from \cite{dainotti2013} to a larger sample of LGRBs and a sample of SGRBs, which were shown to likely to follow the same correlation \citep{rowlinson2013}. In Section 3, we derive the intrinsic LT correlation slope and normalisation using the magnetar central engine model and in Section 4, using fully simulated datasets, we confirm that the observed LT correlation is an expected consequence of the magnetar central engine model. We then use the correlation to place constraints on the magnetar emission beaming and efficiency properties. Finally, Section 5 discusses the magnetar model in the context of the other postulated origins for the LT correlation.

\section{The observed LT Correlation}

\cite{dainotti2008, dainotti2010}, using the \cite{willingale2007} phenomenological law for the lightcurves of GRBs, discovered a formal anti-correlation  between the X-ray luminosity at the end of the plateau $L_{\rm X}$ and the rest-frame plateau end time, $T_{\rm a}^{*}=T_{\rm a}/(1+z)$, described as:

\begin{eqnarray}
\log L_{\rm X} = a + b \log T^{*}_{\rm a} \Leftrightarrow L_{\rm X} = AT_{\rm a}^{*b},
\label{feq}
\end{eqnarray}
where $T^{*}_{\rm a}$ is in seconds, $L_{\rm X}$ is in erg s$^{-1}$ and $z$ is the redshift. The normalisation and the slope parameters ($a$ and $b$) are constants obtained by the D'Agostini fitting method \citep{dagostini2005}, which employs a Bayesian approach to probabilistically infer the best fitting linear model for data with both statistical uncertainties and an intrinsic scatter about the correlation. \cite{dainotti2013} demonstrated through the \cite{efron1992} method, a statistical method to determine that a correlation between two parameters exists when the dataset suffers from observational biases (such as selection effects caused by redshift evolution), that the LT correlation is not caused by observational selection effects, but is an intrinsic correlation at a 12$\sigma$ level. The strategy implemented in \cite{efron1992} does not retain the normalisation of the correlation and, hence, we are unable to determine the intrinsic normalisation of the correlation ($a_{\rm int}$). \cite{dainotti2013} showed that the intrinsic slope is $b_{\rm int}=-1.07_{-0.14}^{+0.09}$. Therefore, as it is intrinsic to the source and not caused by selection effects, the LT correlation is a useful tool for the testing of theoretical GRB physical models \citep[e.g.][]{ghisellini2008, cannizzo2009, yamazaki2009, dallosso2011, cannizzo2011, bernardini2012, rowlinson2013}.

In this paper, we compare the intrinsic correlation to the analytic prediction from the magnetar central engine model and create simulated datasets (as described in Section 4) to investigate the observed LT correlation. The simulated datasets take into account cosmological and observational selection effects, such as plateaus that are undetectable due to sensitivity limits caused by a combination of their luminosities, durations and the redshift distribution. We are simulating datasets, which can be directly compared to the observed LT correlation \citep[without requiring the method employed by][to obtain the intrinsic LT correlation]{dainotti2013} allowing us to retain the normalisation of the correlation. However, the model predicts bolometric luminosities, so we need to compare the simulations to a bolometric approximation of the observed luminosities (using an energy band of 1--10000 keV). Hence, we compare the magnetar model predictions to two different values of the slope of the correlation (the intrinsic slope, $b_{\rm int}$, for comparison to the analytical predictions in Section 3 and the observed slope, $b_{\rm obs}$, for comparison to the simulations in Section 4) and an observed normalisation ($a_{\rm obs}$) which is compared to both the analytic predictions in Section 3 and the simulations in Section 4. In order to obtain the required parameters, in the 1-10000 keV energy band, we refitted the LT correlation. As we refit the data, we also took the opportunity to increase the GRB sample size to reduce the statistical uncertainties utilised in Section 4. In the remainder of this Section, we describe the method utilised to obtain the observed 1--10000 keV LT correlation.

Our sample constitutes all GRB X-ray afterglows with known redshifts detected by {\it Swift} \citep{gehrels2004} from January 2005 up to December 2013, for which the light curves include early X-Ray Telescope \citep[XRT;][]{burrows2004} data and therefore can be fitted using the \cite{willingale2007} phenomenological model. The total sample has been subdivided into 3 categories: LGRBs (149 of the GRBs are in this sample), EE SGRBs (2 GRBs) and SGRBs (8 GRBs). SGRBs are separated from the LGRB sample using the standard T$_{90}\le2$ s cut \citep{kouveliotou1993} to remain consistent with the results presented in \cite{rowlinson2013}\footnote{Note there is debate regarding the true division in T$_{90}$ duration between LGRBs and SGRBs, for instance \cite{bromberg2013}}. The EE SGRBs are from the sample fitted to the magnetar model by \cite{gompertz2013}, which also have a redshift and can also be fitted using the  \cite{willingale2007} phenomenological model. We use the redshifts available in the literature \citep[][and references therein]{xiao2009} and redshifts taken from GRB Coordinates Network circulars (GCNs)\footnote{Utilising http://www.mpe.mpg.de/$\sim$jcg/grbgen.html}. We exclude all GRBs with non-spectroscopic redshifts. As the majority of SGRBs do not have observed redshifts, we use those with a firm host galaxy association \citep[from the sample defined in][]{rowlinson2013}. The complete sample of GRBs analysed contains 159 events, covering the redshift range $0.033 \leq z \leq 9.4$. In our analysis, we adopt a flat cosmology with $H_0 = 69.9$ km s$^{-1}$ Mpc$^{-1}$, $\Omega_{\rm M} = 0.28$ and $\Omega_{\lambda} = 0.72$ \citep[ see][for a detailed discussion regarding different cosmological models]{dainotti2013}.

We note that this sample is larger than that used by \cite{dainotti2013} to identify the intrinsic correlation ($b_{\rm int}$). We do not recalculate the intrinsic correlation as the distributions of plateau durations, fluxes and spectral indices remain the same as those utilised in \cite{dainotti2013b}, so the GRB populations are directly comparable for this purpose. Additionally, the limiting fluxes and plateau durations are also unchanged for this sample of GRBs. As the only significant difference is the sample size, we are confident that this will not significantly change the intrinsic slope (within 1$\sigma$ uncertainties).

The combined {\it Swift} Burst Alert Telescope \citep[BAT,][]{barthelmy2005} and XRT light curves of the GRBs were converted to rest-frame lightcurves using the observed X-ray spectral index for each GRB, a k-correction and the methods described in \cite{bloom2001} and \cite{evans2009}. As we intend to compare the observed distribution to the predictions from a bolometric model \citep[in contrast to ][where an XRT band pass k-correction was used]{dainotti2010,dainotti2013b}, we use an approximate rest-frame bolometric energy band (1--10000 keV). We fitted the lightcurves with a two component model consisting of an initial steep decay phase for the early X-ray emission and an afterglow component \citep[utilising the methods described in][]{willingale2007,dainotti2008,dainotti2010,dainotti2013}. We assume that the rise time of the afterglow component is a free parameter \citep[whereas in][the rise time of the afterglow is assumed to be equal to the start time of the initial decay phase]{willingale2007} so that we can search for an independent measure of the break time. We fitted the lightcurves for which the break time and flux were reliably determined by the model. Previous analyses by \cite{dainotti2008,dainotti2010,dainotti2011b,dainotti2013} utilised the \cite{avni1976} prescription to obtain the required parameters of the plateau (the flux of the plateau, the plateau duration and the decay index following the plateau phase). \cite{avni1976} developed a method to estimate the uncertainty ranges for only the parameters of interest within a fitted model. This method uses the `best-fit' value of the parameters of interest and their corresponding $\chi_{\rm best}^2$. The parameter values are varied until the $\chi^2$ of the fit increases by a particular amount above $\chi_{\rm best}^2$, referred to as the critical $\Delta\chi^2$. $\Delta\chi^2$ depends upon the number of parameters that are estimated simultaneously and not the total number of parameters in the model. The critical $\Delta\chi^2$ is dependent upon the required confidence level (68\% in this analysis) and the number of parameters being varied simultaneously \citep[typical values are given in Table 1 of][]{avni1976}. In \cite{dainotti2008,dainotti2010,dainotti2011b,dainotti2013} the value $\Delta\chi^2<3.5$ was used as they required values for these fitted parameters: plateau flux, plateau duration and the plateau temporal slope. However, in this paper we want to use the largest possible sample of GRBs and we use $\Delta\chi^2<2.3$. This is appropriate as we are only interested in two of the parameters (plateau flux and duration) that are typically fitted in the model and neglect to fit the plateau slope as it does not enter into the computation of the luminosity. The $\chi^{2}$ distribution for some GRBs in the sample is not parabolic out to a value of 3.5 so the \cite{avni1976} prescription is not fulfilled and they are discarded because the evaluation of their error parameters is not precise. However, when the constraint is dropped to 2.3, the $\chi^{2}$ distributions of more of the GRBs in the sample are parabolic and meet the \cite{avni1976} prescription. Hence, this change increased the sample by 20 GRBs which were recovered from the previous sample from January 2005 till March 2013\footnote{The fit has been performed with the package NonlinearModelFit in {\sc Mathematica 9}; the data and the code are available upon request to maria.dainotti@riken.jp}.

\begin{figure}
\centering
\includegraphics[width=0.46\textwidth]{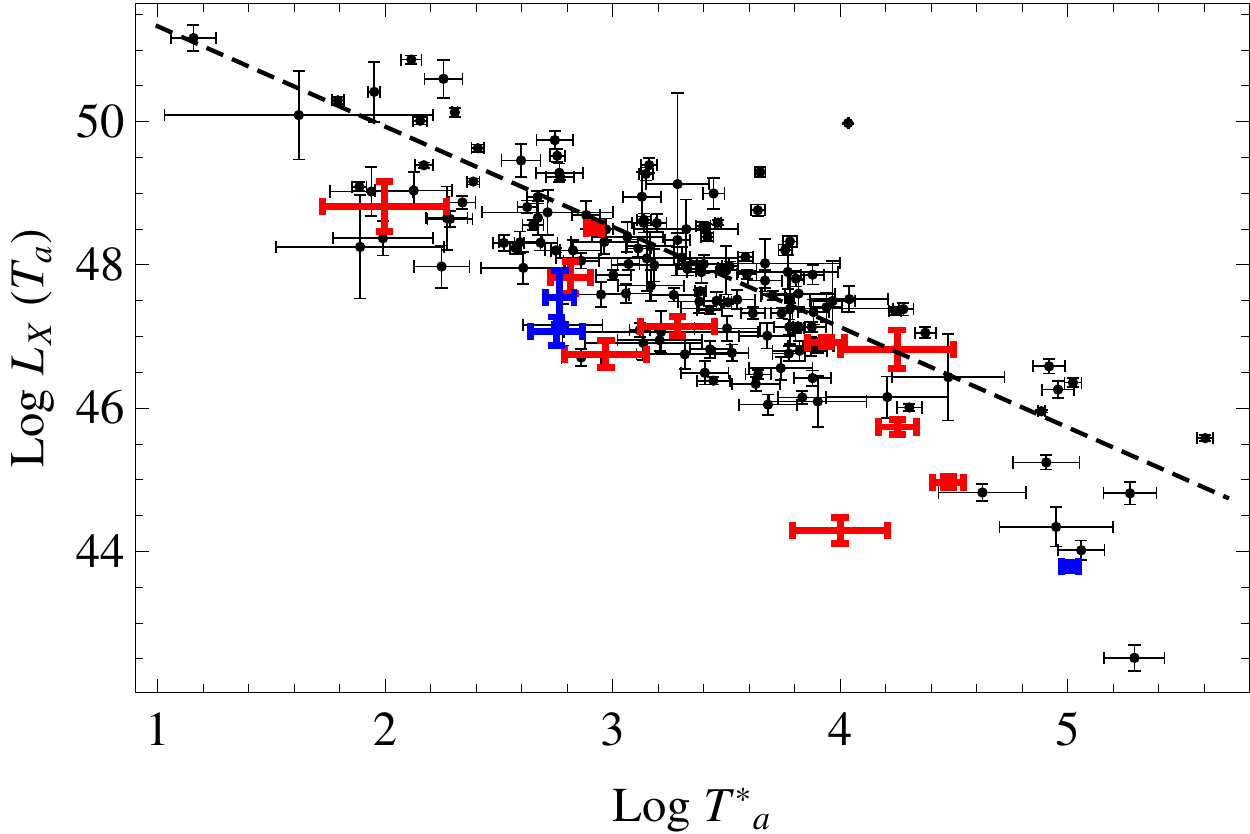}
\caption{The rest-frame plateau durations versus the luminosity (1--10000 keV) at the end of the plateaus for all the GRBs in the sample (black = LGRBs, Blue = EE SGRBs and Red = SGRBs). Over-plotted, using the dashed black line, is the observed LT correlation for the full sample.}
\label{LT_plot}
\end{figure}

From the fitted lightcurves we computed the 1--10000 keV luminosity at the end of the plateau phase and the rest-frame break time. The total sample is fitted with the LT correlation (Equation \ref{feq}) and we find a slope of $b_{\rm obs} = -1.40\pm 0.19$ and a normalisation of $a_{\rm obs} = 52.73\pm0.52$, as shown in Figure \ref{LT_plot}. The data are scattered around this correlation, with a standard deviation of 0.89. These parameters represent the observed correlation, which is found to be steeper than the intrinsic correlation \citep[due to redshift dependencies as discussed in][]{dainotti2013}. The redshift dependences are instead accounted for within the modelling used to simulate the correlation (as described in Section 4). We note that the SGRBs and EE SGRBs typically are offset from the observed correlation suggesting that, although they appear to follow the same correlation, they may have a different normalisation. This may be associated with different redshift distributions (and hence observational constraints) or different beaming/efficiencies as we describe in Section 4. By conducting a multi-dimensional Kolmogorov-Smirnov test \citep[KS test;][]{harrison2014,metchev2002,gosset1987}, we can test if the SGRBs and EE SGRBs are being drawn from the same distribution as the LGRBs. We applied a multi-dimensional KS test for the distributions of the durations, luminosities and their associated errors ($\log T^{*}_{a}$, $\delta \log T^{*}_{a}$, $\log L_{X}$ and $\delta \log L_{X}$) for the two samples, LGRBs versus SGRBs and EE SGRBs, and obtain a p-value of $\sim7\times10^{-4}$. Therefore, we can confidently conclude that the SGRBs and EE SGRBs are drawn from different distribution to the sample of LGRBs. However, as there are only a small number of SGRBs (8) and EE SGRBs (2) in the sample, there are currently insufficient data to be able to make significant quantitative comparisons between the different categories of GRBs.

\section{The magnetar model and LT correlation}

A newly formed magnetar, predicted to form via a range of mechanisms such as accretion induced collapse of a white dwarf, collapse of a massive star and the merger of two NSs \citep[e.g.][]{usov1992,thompson1994,dai1998}, will have a vast supply of rotational energy that will be emitted via dipole radiation \citep{zhang2001}. Initially the dipole radiation is expected to power a plateau phase which ends when the characteristic dipole spindown timescale is reached \citep[as defined in][]{zhang2001}. In the magnetar model, the energy released by dipole radiation is predicted to be transferred to the surroundings via a magnetar wind \citep{metzger2011}. \cite{dallosso2011} considered the specific case where the magnetar wind refreshes the forward shock emission and \cite{bernardini2012} applied this interpretation to show that the slope of LT correlation is consistent with the magnetar central engine model (c.f. equation B.3). However, there are a number of both LGRBs and SGRBs exhibiting a steep decay phase following the plateau phase which are inconsistent with the forward shock model \citep{troja2007,lyons2010,rowlinson2010,rowlinson2013}. In these cases, the plateau is interpreted as being an extra component (internal to the central engine) in addition to the standard forward shock afterglow \citep{troja2007}. In the SGRB sample $\sim$29--56\% of the magnetar candidates have a steep decay phase \citep{rowlinson2013}, whereas this is likely to be somewhat lower for the LGRB case \citep[10 candidates identified to date;][]{lyons2010}. In the specific case of SGRBs, they are also expected to occur in low density environments due to large offsets from their host galaxies \citep[e.g.][]{fox2005,gehrels2005,hjorth2005,fong2010,church2011,tunnicliffe2014} so they are expected to have little material for the forward shock to interact with, giving a much fainter afterglow component. Therefore, the refreshed forward shock model is difficult to reconcile with the full sample of GRBs fitted with the magnetar model. Hence, in the following analysis, we have assumed a simple model of the energy released by dipole radiation from a spinning down magnetar and do not consider the mechanism by which this energy is transferred to the observed emission. This enables us to test the intrinsic relationship between the magnetar central engine model and the LT correlation. The bolometric luminosity and plateau duration are intrinsically associated and can be calculated, to a reasonable approximation, using these Equations \citep{zhang2001}:

\begin{eqnarray}
L_{0,49}\sim(B^2_{{\rm p},15}P^{-4}_{0,-3}R^6_6)\label{luminosity}\\
T_{{\rm em},3}=2.05~(I_{45}B^{-2}_{p,15}P^2_{0,-3}R^{-6}_6),\label{duration}
\end{eqnarray}
where $T_{{\rm em},3}$ is the rest-frame plateau duration in $10^{3}$ s, $L_{0,49}$ is the plateau luminosity in $10^{49}$ erg s$^{-1}$, $I_{45}$ is the moment of inertia in units of $10^{45}$g cm$^{2}$, $B_{{\rm p}, 15}$ is the magnetic field strength at the poles in units of $10^{15} G$, $R_{6}$ is the radius of the neutron star in $10^{6}$ cm and $P_{0,-3}$ is the initial period of the compact object in milliseconds. Using these equations, we derive an intrinsic correlation between the plateau luminosity and durations:
\begin{eqnarray}
\log(L_{0})\sim \log(10^{52} I_{45}^{-1} P^{-2}_{0,-3}) - \log(T_{{\rm em}}) \label{intrinsic2}
\end{eqnarray}

Therefore, there is an intrinsic correlation expected from the magnetar model where $L \propto T^{-1}$, with some scatter dependent on the range of initial spin periods and moment of inertia of the NS. By comparing this to Equation \ref{feq} we show $b_{\rm predict}=-1$, which is consistent (to within the 1$\sigma$ uncertainties) with the intrinsic LT correlation, $b_{\rm int}=-1.07^{+0.09}_{-0.14}$. Additionally, we show that the normalisation to the predicted LT correlation is given by $a_{\rm predict}=\log(10^{52} I_{45}^{-1} P^{-2}_{0,-3}) \sim 52$ (assuming $I_{45}=P_{0,-3}=1$), which is also comparable to the observed 1--10000 keV normalisation $a_{obs}=52.73\pm0.52$ (Section 2).

In this analysis we have assumed 100\% efficiency in the transfer of rotational energy to electromagnetic radiation and that it is isotropically emitted, which is likely to be an unphysical assumption. This assumption is expected to change the normalisation value but not to change significantly the intrinsic slope, because there is no dependence on the luminosity for $b_{\rm predict}$. However, as shown in Section 4, there are observing constraints limiting the parameter space that can be probed and therefore strongly affect the slope of the correlation ($b_{\rm sim}$ and $b_{\rm obs}$). The dependence on beaming and efficiency are given by $L \propto \frac{\epsilon}{1-\cos\theta}$, so we can derive the corrected normalisation:

\begin{eqnarray}
a_{\rm corr} = \log(\epsilon) - \log(1-\cos\theta) + a, \label{normalisation}
\end{eqnarray} 
where $\epsilon$ is the efficiency (which describes the proportion of the spindown luminosity of the magnetar that has been transferred to observable radiation) and $\theta$ is the beam opening angle in degrees. Therefore, we can use the observed normalisation value to constrain the, currently unknown, beaming angle and efficiency of the magnetar dipole emission. This further extends the analysis of \cite{bernardini2012}, as the efficiency was a fitted parameter in their model, with high efficiencies ranging within 25--99\% and an average efficiency of $\sim70$\%. With a reasonable range of efficiencies (1 -- 100 \%) and beaming angles (isotropic -- 1 degree), we predict the normalisation will range between $(a - 3.8)$ and $(a + 2.4)$  corresponding to $a_{\rm corr}$=48.2 -- 54.4, a range of values significantly larger the observed uncertainties. This enables us to use the observed normalisation, $a_{\rm obs}$, to constrain the beaming and efficiency of the magnetar central engine.

Hence, analytically, we have shown that an LT correlation is intrinsic to the spindown of a newly formed magnetar \citep[confirming the findings by][]{bernardini2012} and is consistent with the intrinsic LT correlation, $b_{\rm int}$, observed by \cite{dainotti2013}. Additionally, we have shown the predicted normalisation, $a_{\rm predict}$, directly depends on the beaming angles and efficiency of the magnetar emission. Hence, we can use the observed 1--10000 keV normalisation, $a_{\rm obs}$, to constrain the beaming angle and efficiency. In the next Section, we confirm that the observed correlation ($b_{\rm obs}$) can be obtained from a simulated dataset, which takes into account observational biases for the magnetar central engine model, and use the simulations with $a_{\rm obs}$ to constrain the beaming and efficiency of the magnetar emission.

\section{Simulating the LT correlation using the magnetar central engine model}

\begin{figure}
\centering
\includegraphics[width=0.46\textwidth]{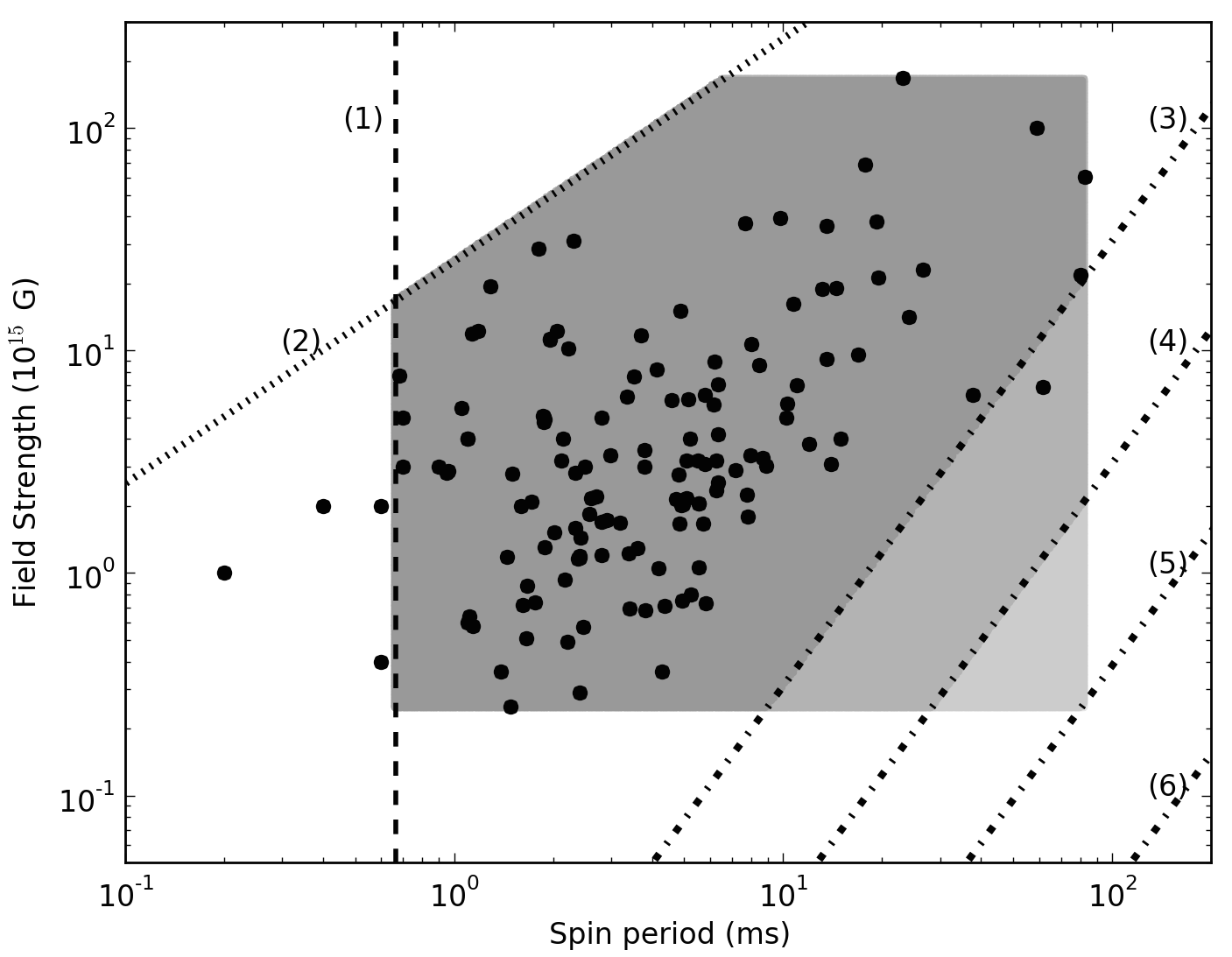}
\caption{The grey shaded regions represent the uniform distribution of values of magnetic field strengths and spin periods used to simulate the observable magnetar plateaus. The upper and lower limits on the magnetic field strength and the upper limit on the spin period are determined using the sample of GRBs fitted with the magnetar model \citep[over-plotted as black circles;][]{lyons2010,dallosso2011,bernardini2012,gompertz2013,rowlinson2013,postigo2013,yi2014,lu2014}. The dashed black vertical line (1) at 0.66 ms represents the minimum spin period allowed before break-up of a 2.1 M$_{\odot}$ neutron star. The dotted black line (2) represents a limit on spin periods and magnetic field strengths imposed by the fastest slew time of XRT in our sample in the rest-frame of the highest redshift GRB in the sample, as plateaus with durations shorter than the slew time are unobservable. The black dash-dot lines (3--6) represent the observational cut-offs for the faintest XRT plateau observable assuming the lowest redshift in the GRB sample. These cut-offs change depending on the beaming and efficiency of the magnetar emission: (3) Minimum beaming angle and efficiency (1 degree and 1\% respectively), (4) Minimum efficiency (1\%) and maximum beaming angle (isotropic), (5) Maximum efficiency (100\%) and minimum beaming angle, (6) Maximum efficiency and beaming angle. The observed distribution follows the region bounded by lines (1--3) suggesting that the sample have low efficiencies and small beaming angles, as otherwise candidates would have been observed in this region.}
\label{BP_plot}
\end{figure}

In the previous Section, we showed that an LT correlation is a natural consequence of the magnetar model, with a scatter that is determined by the range of allowed spin periods and inertias. To confirm that the observed correlation ($a_{\rm obs}$ and $b_{\rm obs}$) can be obtained from the magnetar model, we create simulated datasets based on different values of the magnetic field strength and spin period (assuming a fixed inertia and NS radius) that exclude reasonable regions of parameter space due to observing constraints and understanding of the underlying model. This strategy significantly differs from the approach of \cite{bernardini2012} because we are using predictions of the properties of a newly formed magnetar and considering observational selection effects, whereas they drew their sample from a Gaussian distribution about the mean properties fitted to 12 GRBs. By using fully simulated values, we ensure that we are accounting for the entire parameter space when we derive the LT correlation. We illustrate all the constraints on the parameter space in Figure \ref{BP_plot}. 

We place constraints on the spin period using a minimum spin period, equal to spin break-up, calculated using:
\begin{eqnarray}
P_{0,-3} \ge 0.81 M_{1.4}^{-1/2}R_{6}^{3/2} {\rm ms}, \label{minP}
\end{eqnarray}
(assuming a mass of 2.1 M$_{\odot}$) and illustrated as the dashed line (1) bounding the left hand side of the shaded region in Figure \ref{BP_plot}.
We use the maximum and minimum values from fits of the magnetar model to observed GRB lightcurves to set the maximum spin period and the range of magnetic fields to simulate \citep[black dots in Figure \ref{BP_plot}, from the samples of ][]{lyons2010,dallosso2011,bernardini2012,gompertz2013,rowlinson2013,postigo2013,yi2014,lu2014}, giving $P_{\rm max}\sim83$ s, $B_{\rm max}\sim 2\times10^{17}$ G and $B_{\rm min}\sim 3\times10^{14}$ G. These limits define the upper, lower and right hand bounds on the shaded region in Figure \ref{BP_plot}. We note an important caveat that these samples have all been fitted using different assumptions on beaming angles and efficiencies and, as shown in \cite{rowlinson2010, rowlinson2013}, this can strongly affect their magnetic field strengths and spin periods. There are a number of candidates in the sample whose maximum spin periods are in excess of the suggested upper limit on the birth spin period \citep[10 ms;][]{usov1992}. However, this is unsurprising as the magnetar may have undergone initial rapid spin down due to gravitational wave emission \citep[][]{zhang2001} and accretion or propellering \citep[e.g.][]{piro2011,gompertz2013b}. Additionally, there are a small number of candidates whose spin periods are faster than the spin break-up period of a 2.1$M_{\odot}$ NS, which are not physically possible. With different assumptions about beaming and efficiency, these candidates can move into the allowed parameter space \citep[as shown in][]{rowlinson2010,rowlinson2013}.

There are further observational constraints that affect the duration and luminosity of observable plateaus. {\it Swift} is capable of slewing to observe GRBs very rapidly, but this is not instantaneous and if the plateau is shorter than the slew time the plateau would be unobservable. We reject any combinations of magnetic fields and spin periods that would result in a plateau that is shorter than the shortest observable plateau, defined as being the shortest {\it Swift} slew time from the sample in Section 2 ($\sim50$ s) in the rest-frame of the highest redshift GRB in the sample \citep[$z\sim9.4$;][]{cucchiara2011}. This limit is represented by the dotted line (2) in Figure \ref{BP_plot}. The second observational constraint is the lowest luminosity plateau that {\it Swift} XRT would be able to detect, corresponding to the lowest flux detectable by XRT \citep[$2\times10^{-14}$ erg cm$^{-2}$ s$^{-1}$;][]{burrows2004} for the nearest GRB in the sample \citep[$z\sim0.03$;][]{mirabal2006}. The lowest luminosity magnetar plateau that would be observable by {\it Swift} is dependent on the beaming angle and efficiency, so we vary this limit accordingly and obtain the observable combinations of $B_{{\rm p},15}$ and $P_{0,-3}$ using Equation \ref{luminosity}. We use beaming angles in the range $1^{\circ}$ -- $90^{\circ}$ and efficiencies from 1\% to 100\%. The extremities of these limits are illustrated using the dash-dotted lines (3--6) in Figure \ref{BP_plot}. Due to the observational constraints, we might expect to observe slightly different correlations for the sub-samples of SGRBs, EE SGRBs and LGRBs due to their different redshift distributions. Additionally, changes in the normalisation of the correlation could be associated with different beaming or efficiencies for these populations, which are discussed further in this Section. The locations of SGRBs and EE SGRBs in Figure \ref{LT_plot} suggest that this may indeed be occurring.

It is clear from Figure \ref{BP_plot} that the observable distribution of magnetic fields and spin periods is controlled by selection effects. In particular, we do not expect to observe high magnetic fields and rapid spin periods. Secondly, {\it Swift} is unlikely to detect magnetars with long spin periods and low magnetic fields. The distribution of the fitted GRBs in Figure \ref{BP_plot} is consistent with the emission from their magnetar central engines being beamed and of low efficiency.

Using the regions defined in Figure \ref{BP_plot}, we simulate a large range of magnetic fields and spin periods (uniformally filling out the defined parameter space with $2.5\times10^{5}$ different combinations in logarithmic bins) and randomly sample 159 (equal to the number of GRBs used in Section 2) combinations of magnetic field strength and spin period. The plateau and luminosity of each combination is calculated using Equations \ref{luminosity} and \ref{duration}. The simulated luminosities and durations are fitted using the LT correlation (Equation \ref{feq}). We repeat this random sampling 500 times and combine them to calculate the mean and standard deviation for the slope and normalisation. This simulation is repeated for $2.5\times10^5$ different combinations of beaming angles (1$^{\circ}$--90$^{\circ}$) and efficiencies (1--100\%), evenly distributed across the parameter space in logarithmic bins for the efficiencies and linear bins for the beaming angles. Finally, as the slope of the correlation is not expected to vary significantly (and was confirmed to only fluctuate within the 1$\sigma$ uncertainties), we calculate the average slope for all the simulations and find $b_{\rm sim}= -1.30\pm0.03 $. This is steeper than the intrinsic and analytically predicted slope, $b_{\rm int}$ and $b_{\rm predict}$, which is caused by the observational selection effects that were included in the simulated datasets. However, the simulated slope is well within the $1\sigma$ uncertainties on the observed 1--10000 keV slope ($b_{\rm obs}=-1.40\pm0.19$, as given in Section 2) that did not account for selection effects, showing that the two correlations (simulated and observed) are consistent with each other. We note that there is some low significance variation ($< 1\sigma$) in the fitted slope that correlates with the beaming and efficiency values, with a best match between the observed and simulated slopes being at low efficiencies and beaming angles.

As shown in Section 3, the simulated normalisations ($a_{\rm sim}$) are highly dependent upon the beaming angle and efficiencies used. To quantify the quality of the simulated models in comparison to the observed dataset, we calculate the $\chi^2$ (with 157 degrees of freedom) between the model and observed data. We calculate the $\chi^2$ using:
\begin{eqnarray}
\chi^2 = \sum (y_i - bx_i - a)^2, \label{chisq}
\end{eqnarray}
where the observed dataset from Section 2 is inserted as the values for $x_i$ and $y_i$, while $a$ and $b$ are from the model being tested. We then use the $\chi^2$ and degrees of freedom to calculate the probability that the observed data are consistent with being drawn from the model obtained from the simulation. Models that are offset from the observed data, due to having different normalisations, will have a much lower probability and are therefore inconsistent with the data.

\begin{figure}
\centering
\includegraphics[width=0.46\textwidth]{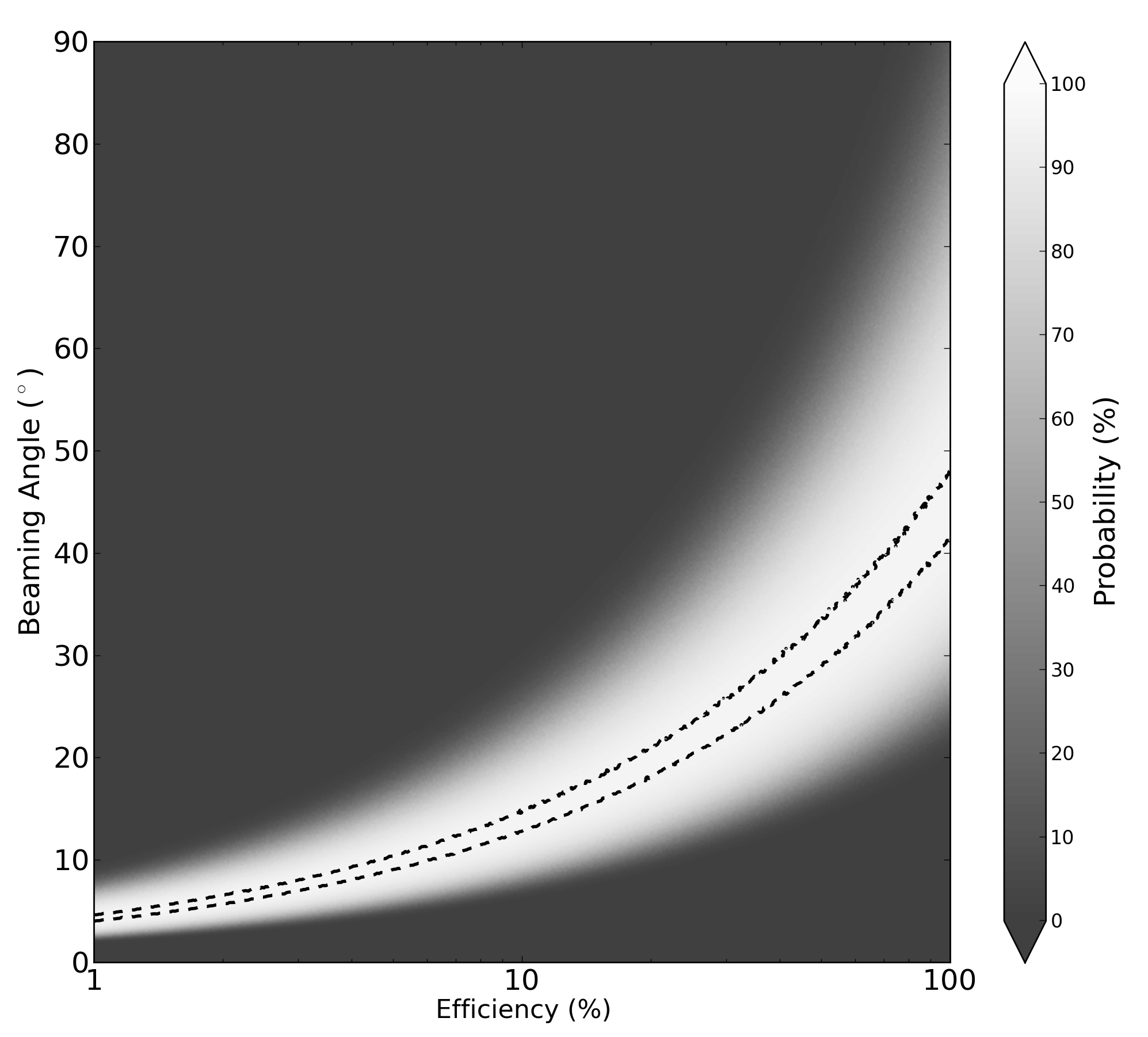}
\caption{Simulated datasets were created with a wide range of efficiencies (1--100\%) and beaming angles (1--90$^{\circ}$), which were then fitted with an LT correlation. This figure plots the probability that the observed data can be drawn from the simulated model, calculated using the $\chi^2$ (as defined by equation \ref{chisq}). The dashed lines bound the region where the simulated LT correlation has $>$95\% probability of explaining the observed data. The darkest region shows where the simulated LT correlations have $<$10\% probability of explaining the observed data.}
\label{beam_eff}
\end{figure}

In Figure \ref{beam_eff}, we show a contour plot of the probability that the data are drawn from the model, as a function of the modelled beaming angles and efficiencies. The darkest grey regions are for those models which have a $< 10$\% probability that the data can be drawn from them. The black dashed contour lines show boundary within which the data have a $> 95$\% probability of being drawn from the simulated models. This parameter space is a region sweeping from the lowest efficiencies and beaming angles of $\sim5^{\circ}$ up to the highest efficiencies and beaming angles of $\sim40-50^{\circ}$. It seems unlikely that the conversion of rotational energy to the observed X-ray emission is highly efficient.

The efficiency of converting energy in the GRB blast wave to gamma-ray emission, has been well studied both theoretically and observationally with typical values lying in the range 1--10\% with a narrow spread of values between different GRBs \citep[e.g.][]{freedman2001,nysewander2009,davanzo2012} and SGRBs are thought to have comparable efficiencies to LGRBs \citep{berger2007, nakar2007}. Additionally, both LGRBs and SGRBs are expected to be beamed with typical beaming angles being $<20^{\circ}$ \citep[from both simulations and observations, e.g.][]{berger2003,zhang2006b,racusin2009,fong2014}, however SGRBs are expected to be less collimated than LGRBs \citep[e.g.][]{janka2006,berger2007}. These low efficiencies and narrow beaming angles are all fully consistent with the highest probability region shown in Figure \ref{beam_eff}. \cite{bucciantini2009} simulated the jets produced by a magnetar central engine in the context of a collapsar progenitor and determined that collimation is facilitated by the stellar envelope giving typical beaming angles of 5--10$^{\circ}$. Using Figure \ref{beam_eff}, this would correspond to an efficiency of $\sim$5--10\% which is also consistent with the expected efficiencies. Additionally, as collimation is facilitated by the stellar envelope, in the context of the binary neutron star progenitor model it is reasonable to expect the beam to have a wider beaming angle as there is minimal surrounding material. \cite{metzger2011} studied the emission mechanisms from magnetars, typically assuming a radiative efficiency of 50\%, and determined that a beaming angle of $\sim$30$^{\circ}$ is most consistent with the late time emission from the magnetar winds. These parameters lie within the 95\% confidence region in Figure \ref{beam_eff}. Therefore the beaming and efficiency values obtained using the simulations presented in this paper, in comparison to the observed dataset, are fully consistent with both the predicted and observational constraints for GRBs and with the predictions in the specific case of the magnetar central engine models.

\begin{figure}
\centering
\includegraphics[width=0.46\textwidth]{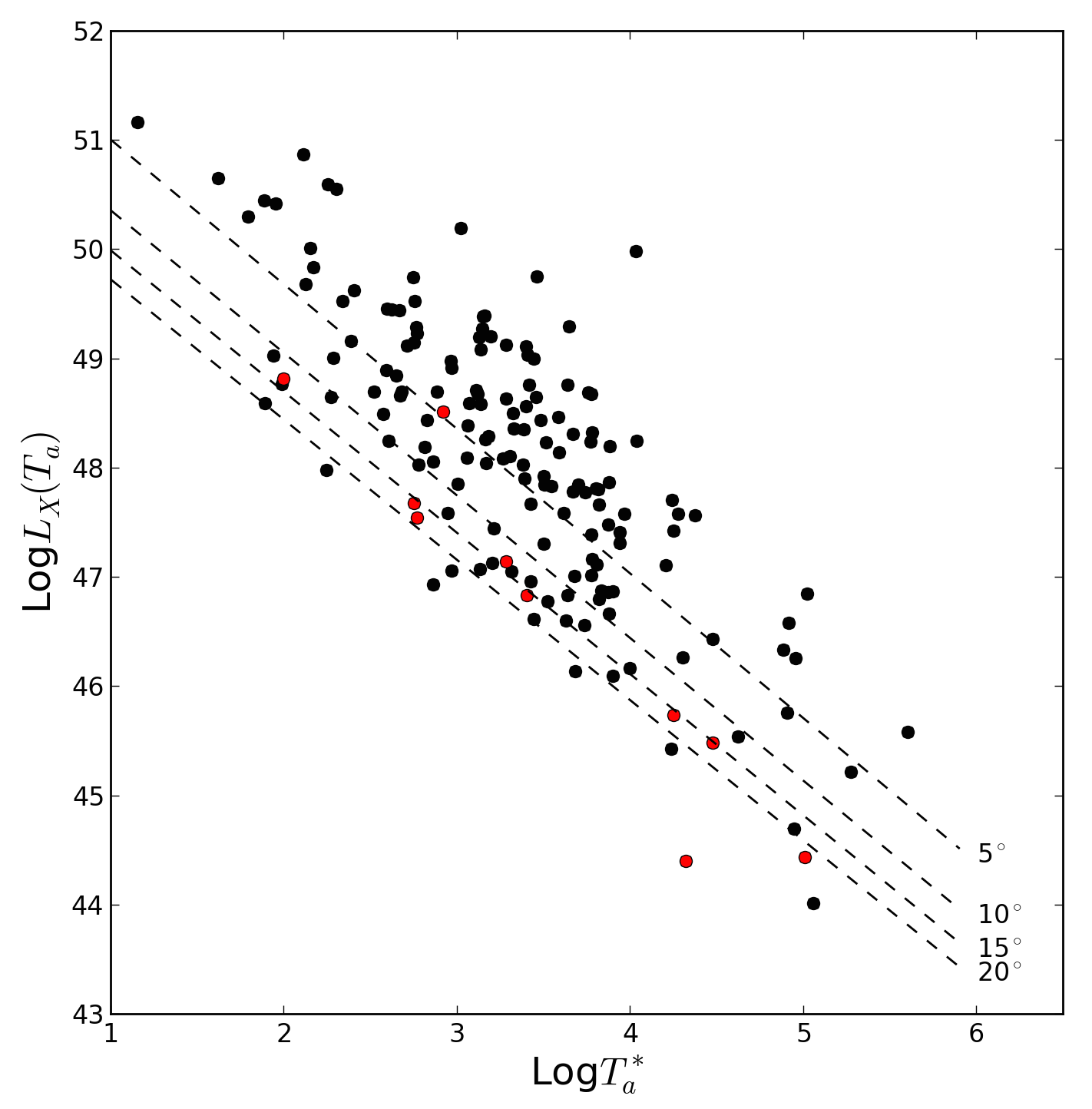}
\caption{This plot shows the effect of increasing the beaming angle (5--20$^{\circ}$) for a given efficiency (1\%). The black data points are the LGRBs from Figure \ref{LT_plot} and they are scattered around the 5$^{\circ}$ line represents a simulation that has a $>$95\% probability of being drawn from the same sample as the observed data. The red data points are both the SGRBs and the EE SGRBs, which are appear to be more consistent with being scattered around the 15$^{\circ}$ line.}
\label{beam_angles}
\end{figure}

In Figure \ref{LT_plot}, there were hints that SGRBs and EE SGRBs may have a lower normalisation than the LGRBs, and a multi-dimensional KS test confirmed that they were likely to be drawn from a different population. If these GRBs do follow a different normalisation, this could be due to different selection effects (e.g. SGRBs tend to occur at lower redshifts) although this is likely to only have a small effect on the simulated sample. Alternatively, it could be caused by differences in the beaming and efficiencies. It has been shown that SGRBs typically have the comparable efficiencies to LGRBs but are expected to have wider beaming angles \citep[e.g.][]{janka2006,berger2007,nakar2007}. In Figure \ref{beam_angles}, we show the effect on the LT correlation of increasing the beaming angle from 5$^{\circ}$ to 20$^{\circ}$, assuming an efficiency of 1\%. The beaming angle and normalisation are clearly inversely proportional to each other and, assuming LGRBs are best fitted with a beaming angle of 5$^{\circ}$ and an efficiency of 1\%, the SGRB population appear to be more consistent with being scattered around a larger beaming angle of $\sim$15$^{\circ}$. This result supports the expectation that SGRBs have wider beaming angles than LGRBs.

\begin{figure}
\centering
\includegraphics[width=0.46\textwidth]{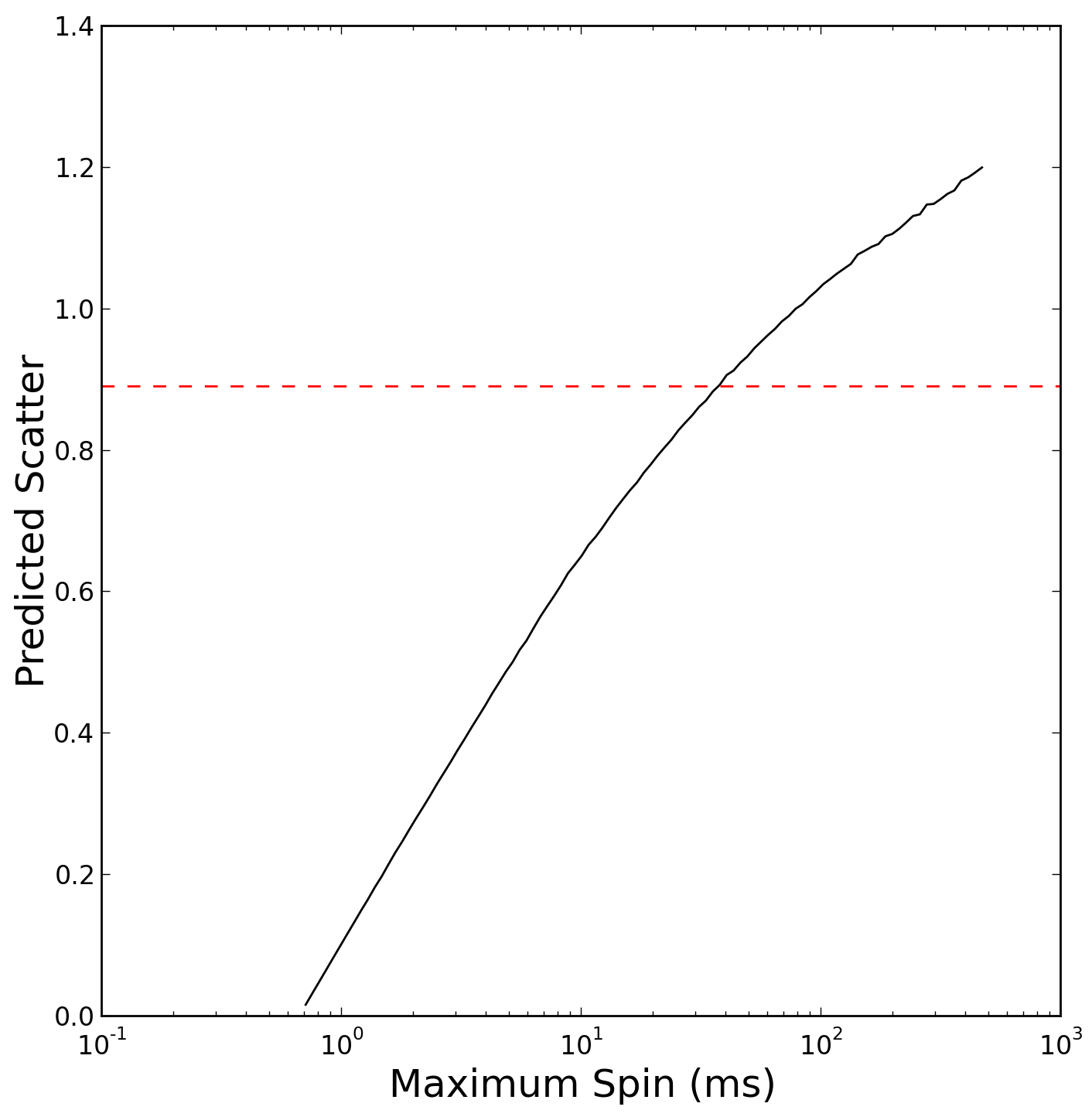}
\caption{In Equation \ref{intrinsic2}, we showed that the scatter of the predicted LT correlation is dependent upon the range of initial spin periods. The black line shows how the predicted scatter of the LT correlation increases as the range of spin periods increases. The range is defined as: spin break up period (0.66 ms, assuming a 2.1 M$_{\odot}$ neutron star) -- maximum spin period. The red dashed line represents the observed scatter of the LT correlation, 0.89, fitted in Section 2. By measuring the intercept of the model and the observed value, the scatter on the observed LT correlation corresponds to a spin period range of 0.66 -- $\sim$35 ms.}
\label{scatters}
\end{figure}

The observed scatter of the correlation was shown to be dependent upon the range of initial spin periods and inertias of the magnetars (Equation \ref{intrinsic2}). The inertias are expected to be in the range $1 \ge I_{45} \ge 1.5$, corresponding to standard neutron stars in the mass range 1.4--2.1 M$_{\odot}$, whereas the spin periods are expected to vary over $\sim$2 orders of magnitude. Therefore, as the scatter is $\propto I_{45}^{-1} P_{0,-3}^{-2}$, the range of spin periods will dominate the observed scatter. As there is a firm lower limit on the spin period, i.e. the spin break up period of a 2.1 M$_{\odot}$ neutron star, we model the range of spin periods by varying the maximum initial spin period. We measure the scatter on the simulated LT correlation for 100 different maximum spin periods logarithmically distributed in the range $0.66 < P_{0,-3, {\rm max}} \le 500$ ms. For each maximum spin period, we simulate and fit 500 randomly sampled distributions. The scatter around each fitted sample is measured and we calculate the average scatter for each of the maximum spin periods. We plot the results in Figure \ref{scatters} and show that the observed scatter, 0.89, is consistent with a sample of magnetars with spin periods in the range 0.66 -- $\sim$35 ms. This is consistent with 97\% of the observed magnetar fits shown in Figure \ref{BP_plot}. It is important to note that this assumes that all the magnetars have identical beaming angles and efficiencies. If the beaming angles and efficiencies vary significantly between different magnetars, it may increase the observed scatter of the predicted correlation.

\section{Discussion}

The LT correlation \citep{dainotti2008,dainotti2010,dainotti2013} is naturally explained by the relationship between $L_{0,49}$ and $T_{{\rm em},3}$ (Equations \ref{luminosity} and \ref{duration}), but there are a number of other models that have been proposed as GRB central engines. The external disc model \citep{cannizzo2009} suggests that the plateau is the result of a transient disc formed of fall-back material, with the plateau luminosity roughly scaling with the mass accumulated during fall back. The authors predict a scaling of $L_{\rm plat} \propto t^{-3/2}_{\rm plat}$, steeper than the intrinsic slope $t^{-1.07^{+0.09}_{-0.14}}$ \citep{dainotti2010}. It has been suggested that the value for the LT correlation is skewed due to a selection bias against fainter plateaus \citep{cannizzo2011}, however these selection biases have been shown to not strongly affect the intrinsic correlation \citep{dainotti2013}.

In the prior emission model \citep{yamazaki2009}, the plateau is created by emission that begins before the prompt spike, but goes undetected until BAT is triggered and XRT is pointed. This emission takes the form $f_0(t) = A_0t^{-\alpha_0}$, where $\alpha_0$ was found to be typically $1.0$ -- $1.5$, consistent with the value in \citet{dainotti2010}. However, there is no physical reason for this consistency: $\alpha_0$ depends on the delay between the initiation of prior emission and the launching of the prompt jet, a quantity which is unlikely to have a physically significant scaling. Since the correlation is intrinsic to the data, the model `acquires' it when it is applied. In addition, \citet{birnbaum2012} placed stringent observational constraints on this model by extrapolating the prior emission backwards to its initial state. They showed that BAT should have been able to detect the prior emission required to explain the plateau emission.

\cite{kumar2008} discuss two separate mechanisms for creating the X-ray plateau in the context of accretion onto a BH. In the first, low viscosity ($\alpha$) allows a substantial ($\sim 1M_{\odot}$) accretion disc to build up, which then accretes uniformly onto the central BH for a duration $t_{acc}$. After this time, the disc is depleted, so $\dot{m}_{\rm BH}$ goes back to tracking $\dot{m}_{\rm fb}$, and a turnover or drop in luminosity is observed. In their formalism, $\dot{m}_{\rm BH} \propto t_{\rm acc}^{-1}$, and $\dot{m}_{\rm BH}$ is constant throughout $t_{\rm acc}$, so that we find $L_{\rm plat} \propto t_{\rm plat}^{-1}$ (with a scatter of $\approx M_{\rm disc}$). This model requires $\alpha < 10^{-2}$, which may not be physical \citep{kumar2008}. The same formalism is true for the second mechanism, only this time the requirement is that the disc is fed by fallback material at the same rate it loses material to accretion onto the central BH (i.e. $\dot{m}_{\rm BH} \sim \dot{m}_{\rm fb}$). If one of these conditions can be satisfied, the model can reproduce the LT relation in the same way that energy injection from the magnetar's dipole spin can.

Interestingly, a reverse shock, powered by energy injection from an arbitrary central engine, can also reproduce the slope of the LT correlation assuming that the X-ray band lies above both the cooling and peak frequencies in the synchrotron spectrum \citep{leventis2014,eerten2014}. \cite{eerten2014b} investigated the LT correlation in the context of a thin shell (standard model with no energy injection) and a thick shell (including arbitrary energy injection). The thin shell model was unable to reproduce the observed LT correlation, however the thick shell model successfully identified an LT correlation for both the forward and reverse shocks using a range of input parameters, assuming the spectral breaks lie in specific regimes. Therefore, \cite{eerten2014b} showed that energy injection is required to explain the observed LT correlation irrespective of the properties of the central engine. The magnetar model is one such central engine that could provide the required energy injection for the model and it is interesting that both models can independently reproduce the LT correlation. It is worth noting that \cite{leventis2014} and \cite{eerten2014, eerten2014b} do not attempt to reproduce the normalisation of the correlation as it depends upon a number of their model parameters and the expected distributions of these parameters. Therefore, this model warrants further investigation in the specific context of the magnetar central engine.

\section{Conclusions}

The magnetar central engine model predicts a plateau phase powered by dipole spin down; we have shown that an LT correlation is a natural consequence of this model. With reasonable assumptions for the mass and radius of the magnetar, the analytically predicted slope ($b_{predict}$) is within the $1\sigma$ uncertainties of the intrinsic slope ($b_{int}$). Additionally, the simulated slope ($b_{sim}$) is also well inside the $1\sigma$ uncertainties of the 1--10000 keV observed slope ($b_{obs}$). Finally, the observed scatter of data points around the LT correlation is also expected in the magnetar model and is associated with an intrinsic scatter in the spin periods and inertias of the magnetar. The scatter around the correlation is intrinsic to the magnetar central engine model and is consistent with the observed scatter if the initial spin of the magnetar is in the range 0.66 -- $\sim$35 ms. Other postulated physical origins of the LT correlation are unable to intrinsically return the observed slope of the correlation to this accuracy and often rely upon fitted parameters from the observed population. The reverse shock model can also reproduce the slope of the LT correlation assuming there is ongoing energy injection into the shock front, which is complementary to the magnetar central engine model. Therefore, the LT correlation is consistent with the theory that the observed plateau phases are powered by a magnetar central engine.

By assuming the LT correlation is caused by a magnetar central engine, we are able to use the observed data to place constraints on the likely beaming angles and efficiencies of the magnetar emission. Using a comparison between simulated datasets and the observational data, we conclude that the magnetar emission is most likely to be narrowly beamed ($<20^{\circ}$) with $\lesssim 20$\% efficiency of conversion of rotational energy to observed X-ray emission, fully consistent with the expected values for GRBs. Additionally, there are good theoretical and observational reasons to conclude that SGRBs and EE SGRBs have wider beaming angles than LGRBs, which would lead to a lower normalisation of the correlation. Using the small sample  of observed SGRBs and EE SGRBs, we were able to show with a multi-dimensional KS test that they are drawn from a different population to the LGRBs, confirming that they may indeed have a lower normalisation than the LGRBs.

\section{Acknowledgements}
We thank the anonymous referee for their useful comments that made a significant improvement to this paper. We thank Michal Ostrowski for his valuable comments. MGD thanks ITHES Group at Riken for fruitful discussion. AR, RAMJW \& AJvdH acknowledge support from the European Research Council via Advanced Investigator Grant no. 247295. BPG acknowledges funding from the Science and Technology Funding Council. MGD is grateful to the current grant of the Japan Promotion of Science JSPS N: 25.03786 and the grant from Polish Ministry MNiSW N through the grant N N203 579840. This work makes use of data supplied by the UK {\it Swift} Science Data Centre at the University of Leicester and the {\it Swift} satellite. {\it Swift}, launched in November 2004, is a NASA mission in partnership with the Italian Space Agency and the UK Space Agency. Swift is managed by NASA Goddard. Penn State University controls science and flight operations from the Mission Operations Center in University Park, Pennsylvania. Los Alamos National Laboratory provides gamma-ray imaging analysis.

\end{document}